\documentclass[a4paper,11pt]{article}
\pdfoutput=1 

\usepackage{jheppub} 

\usepackage[T1]{fontenc} 
\usepackage[makeroom]{cancel}
\usepackage{graphicx}
\usepackage{amsmath}
\usepackage{amsfonts}
\usepackage{amssymb}
\usepackage{color}
\usepackage[export]{adjustbox}
\usepackage{graphicx}
\usepackage{dcolumn}
\usepackage{bm}
\usepackage{simplewick}
\usepackage{array}
\usepackage{appendix}
\usepackage{relsize}
\usepackage{subcaption}
\usepackage[sort&compress]{natbib}

\title{\boldmath Correction to the Friedmann equation with Sharma-Mittal entropy: A new perspective on cosmology }

\author[a,1]{Muhammad Naeem}
\author[a,1]{ and Aysha Bibi,\note{Corresponding author.}}

\affiliation[a]{Department of Mathematics, Quaid-I-Azam University, Islamabad, Pakistan}

\emailAdd{mnaeem@math.qau.edu.pk}
\emailAdd{aysha.qau.pak@gmail.com}

\abstract{As the black hole entropy does not obey the area law on it. The area law, given by $S\sim A^\delta\text{or}A^R$, where $\delta$ and $R$ , for $0\leq\delta\leq 1$,  $0<R\leq 1$, indicates the amount of quantum gravitational deformation effects. This shows that gravity and thermodynamics are closely related. Based on this deep connection, we tried to discuss the effects of Sharma-Mittal entropy on the cosmological equations. We also investigated whether the apparent horizon-enclosed universe was consistence with the generalized second law of thermodynamics. Finally, using the gravity emergence scenario, we derived the modified Friedmann equation and compared it with that derived from the first law of thermodynamics in the presence of Sharma-Mittal entropy. It is observed that we can recover the results for standard cosmology when $R,\delta=1$.}
\makeatletter
\gdef\@fpheader{}
\makeatother
\begin{document}
\maketitle
\flushbottom\section{Introduction}
 Scientists have been strongly motivated to study the structure of the COVID-19 virus from various perspectives due to its rapid global spread in 2020. In his work, Barrow \cite{E1} put forward a novel structure for the geometric nature of black hole horizons that was inspired by the fractal patterns found in nature. He suggested that the event horizon could possess intricate geometry at infinitely small scales by envisioning a series of spheres that gradually diminish in size and surround the horizon. As a consequence of this fractal geometry, the black hole's volume is finite, while its area can be either infinite or finite. Due to the potential quantum-gravitational effects of spacetime forum, the entropy of black holes would increase, as the area of the horizon no longer follows the conventional area law.
The justification presented encourages physicists to investigate cosmological and gravitational events using generalised statistical formalisms, such as non-extensive R\'{e}nyi and Tsallis entropies \cite{E2,E3,E4,E5,E6,E7,E8,E9,E10,E11,E12,E13,E14,E15,E16,E17,E18,E19,E20,E21,E22,E23,E24,E25,E26,E27}, which substitute a power-law distribution for the normal probability distribution. According to \cite{E28,E29}, there is a suggestion that power-law probability distributions are consistent with the particle spectra observed in experiments that have power-law tails. Moreover, many researchers \cite{E21,E22,E24,E25,E26} have recently investigated the thermodynamic characteristics of several black holes. They utilized the R\'{e}nyi entropy as a generalized entropy with a single parameter, as described in \cite{E30,E31,E32}. As a result, the thermodynamic features of some black holes have been examined.

In contrast, the Sharma-Mittal (SM) entropy \cite{E32} is a generalized form of entropy that incorporates both R\'{e}nyi and Tsallis entropies, and has yielded interesting findings in the field of cosmology. By utilizing the vacuum energy, this entropy has the potential to explain the current phase of the universe as demonstrated in several studies  \cite{E12,E13,E14,E15,E16,E23}. While non-extensive entropies have been employed to examine the thermodynamic properties of black holes, none have employed the SM entropy. As a result, we are driven to investigate the thermodynamic behavior of black holes as complex, interacting gravitational systems using the SM entropy, which is mathematically expressed as
\begin{equation}\label{1}
  S_{h}^{SM}=\frac{1}{R}\left[(1+\delta S_{T})^{\frac{R}{\delta}}-1\right].
\end{equation}
Here $S_T=\frac{A}{4G}\left(\frac{A}{A_0}\right)$ is the Tsallis entropy, where $A=(4\pi r^2$) and  $A_0$ are the horizon areas \cite{E25,E32}, $R$ and $\delta$ are the free parameters that will be established through comparison with observational data \cite{E31,E32}. As $R\rightarrow 0$ or $R\rightarrow \delta$, the R\'{e}nyi and Tsallis entropies can be retrieved, respectively.

Numerous studies have established a strong correlation between gravity and thermodynamics, indicating that the laws of thermodynamics can be translated into laws of gravity \cite{E33,E34,E43,E44,E45,E46,E47,E48,E49,E50,E51,E52,E53,E54}. It has been confirmed that gravity has a thermodynamic basis, and the field equation of general relativity is simply an equation of state for spacetime. When viewed as a thermodynamic system, the laws of thermodynamics on large scales can be expressed in terms of gravity. The concept of "gravity thermodynamics" enables one to reformulate the Friedmann equations in terms of the first law of thermodynamics on the apparent horizon and vice versa \cite{E55,E56,E57,E58,E59}. In his quest to comprehend gravity, Padmanabhan \cite{E60} suggested that the expansion of the universe is the result of the emergence of space, which he derived from the difference in the number of degrees of freedom between the bulk and boundary equated to the change in volume, leading to the Friedmann equation that describes the evolution of the universe \cite{E60}. The idea of emerging spacetime has also been applied to Gauss-Bonnet, Lovelock, and braneworld scenarios \cite{E61,E62,E63,E64,E65}.

The objective of this study is to formulate the cosmological field equations for Friedmann-Robertson-Walker(FRW) universe with curvature, utilizing the entropy expression associated with the apparent horizon. The methodology differs from \cite{E33} that of a previous study where the author incorporated the Barrow entropy in the Friedmann equations assuming a flat FRW universe, and the energy flux across the apparent horizon remains constant while the horizon's radius remains unchanged for a short duration. In this research, the first law of thermodynamics is applied to the apparent horizon in the form of $dE = T dS + W dV$, where $dE$ denotes the change in energy inside the apparent horizon resulting from the expansion of the universe that causes a change in volume ($dV$). The work done during the expansion is included in the thermodynamic formulation. Unlike the previous study \cite{E33}, the current research considers an FRW universe with curvature and modifies the geometry (gravity) part of the cosmological field equations based on SM entropy, where the entropy expression depends on the system's geometry in different gravity theories like Einstein, Gauss-Bonnet, and $f(\text{R})$ gravities. Any modifications to the entropy should correspond to changes in the geometry and vice versa. The authors used the concept of emergence to derive modified cosmological equations based on Barrow entropy, assuming that the energy density and the number of degrees of freedom in bulk remain constant, but the horizon area and the number of degrees of freedom on the boundary are affected by the entropy change. The constants $k_B$, $c$, and $\hbar$ are assumed to be 1 for simplicity.

The paper's structure is outlined as follows: Section 2 involves the derivation of the Friedmann equations that have been modified with the incorporation of SM entropy, obtained through the application of the first law of thermodynamics at the apparent horizon. In Section 3, the generalization of the second law of thermodynamics for a universe filled with SM entropy is evaluated. Section 4 presents the derivation of the modified Friedmann equations utilizing the cosmic space emergence scenario. The conclusions are presented in the last section.
\section{Modified form of Friedmann Equation using Sharma-Mittal Entropy}
We assume a spatially homogeneous and isotropic universe described by FRW metric. The line element of FRW universe can be written as
\begin{equation}\label{2}
  ds^2=h_{\alpha\beta}dx^{\alpha}dx^{\beta}+\tilde{r}^2(d\theta^2+sin^2\theta d\phi^2),
\end{equation}
where $\tilde{r}=a(t)r$, $a(t)$ is the scale factor, and $x^0=1$, $x^1=r$ and the two dimension metric $h_{\alpha\beta}=\text{diag}(-1,a^2/1-kr^2)$. The curvature parameter $k$ describes the open, closed, and flat universe by taking the values $k=-1, 1, 0$, respectively. The explicit evaluation of the apparent horizon of the FRW universe gives the apparent radius
\begin{equation}\label{3}
  \tilde{r}_A=\frac{1}{\sqrt{H^2+k/a^2}},
\end{equation}
where $H$ is the Hubble parameter which describes the expansion rate of the universe. According to the first and second laws of thermodynamics, the apparent horizon is an appropriate horizon from a thermodynamic perspective \cite{E58,E65}. Using the definition of surface gravity $\kappa$, we can also associate the apparent horizon with a temperature. On the apparent horizon, the temperature is given as
\begin{equation}\label{4}
  T_h=\frac{\kappa}{2\pi}=-\frac{1}{2\pi \tilde{r}_{A}}\left(1-\frac{\dot{\tilde{r}}_A}{2H\tilde{r}_{A}}\right).
\end{equation}
The temperature associated with the apparent horizon is defined by $T = |\kappa|/2\pi$, when the condition of $\tilde{r}_A \leq 2H\tilde{r}A$ is hold, as a negative temperature is not physically acceptable. Within a small time interval of $dt$, it can be assumed that the apparent horizon radius, $\tilde{r}_A$, is much smaller than $2H\tilde{r}_A$, which means there is no change in volume and the temperature is defined as $T = 1/(2\pi\tilde{r}_A)$ \cite{E48}. The connection between the temperature on the apparent horizon and Hawking radiation was established in \cite{E59}, confirming the existence of this temperature, further.

It is assumed that the matter and energy of the universe are composed of a perfect fluid, with an energy-momentum tensor of the following form
\begin{equation}\label{5}
  T_{\alpha\beta}=(\rho+p)u_\alpha u_\beta+pg_{\alpha\beta}.
\end{equation}
Here energy density and pressure are represented by $\rho$ and $p$, respectively. The four velocity vector and metric tensor are represented by $u_\alpha$ and $g_{\alpha\beta}$, respectively. We suggest that the conservation equation $\nabla_\alpha T^{\alpha\beta} = 0$ holds for the total energy content of the Universe, regardless of the dynamic field equations. This means that
\begin{equation}\label{6}
  \dot{\rho}+3H(\rho+p)=0.
\end{equation}
Here Hubble parameter $H$ is defined as $H = \dot{a}/a$, where $a$ is the scale factor of the universe. As the universe expands, it results in a change in volume. The work density($W$) related to this volume change is defined as \cite{E60}
\begin{equation}\label{7}
  W=-\frac{1}{2}T^{\alpha\beta}h_{\alpha\beta}.
\end{equation}
The work density for the FRW background with energy-momentum tensor (\ref{5}) is calculated as
\begin{equation}\label{8}
  W=\frac{1}{2}(\rho-p).
\end{equation}
We assume that the first law of thermodynamics holds on the apparent horizon, taking the form
\begin{equation}\label{9}
  dE=T_{h}dS_{h}+WdV,
\end{equation}
where the total energy ($E$) of the universe contained within the apparent horizon, can be represented as $E=\rho V$. The temperature ($T_h$) and entropy ($S_h$) of the apparent horizon are also associated with this. In comparison to the standard first law of thermodynamics ($dE = T dS-pdV$), the work term ($-pdV$) is replaced by $WdV$ in this equation, except in a pure de Sitter space where $\rho=-p$. In this case, the work term $WdV$ reduces to the standard $-pdV$. By examining the matter and energy contained within a 3-sphere of radius $\tilde{r}_A$, we find that the differential form gives us further insight as
\begin{equation}\label{10}
  dE=4\pi\tilde{r}^{2}_{A}\rho d\tilde{r}_A+\frac{4\pi}{3}\tilde{r}^{3}_{A}\dot{\rho}dt.
\end{equation}
Assuming the volume $\left(V=\frac{4\pi}{3}\tilde{r}_{A}^{3}\right)$ enclosed by a 3-dimensional sphere with an apparent horizon area of $A=4\pi\tilde{r}_{A}^{2}$. Using the Eq. (\ref{6}), we get
\begin{equation}\label{11}
   dE=4\pi\tilde{r}^{2}_{A}\rho d\tilde{r}_A-4\pi H\tilde{r}^{3}_{A}(\rho+p)dt.
\end{equation}
The central idea is to apply the entropy represented by Sharma-Mittal entropy (\ref{1}) to the apparent horizon, with the only modification being the substitution of the apparent horizon radius, $\tilde{r}_A$, for the black hole horizon radius, $r^{+}$. By taking the Sharma-Mittal entropy (\ref{1}) in its differential form, we obtain
\begin{align}\label{12}
\nonumber
  dS_{h}^{SM}&=d\left[\frac{1}{R}(1+\delta S_T)^{R/\delta}-1\right] , \\
   &=\left[\frac{2\delta\pi}{G}\left(\frac{4\pi}{A_0}\right)^{\delta-1}\tilde{r}_{A}^{2\delta-1}+2(R-\delta)\frac{\delta\pi^2}{G^2}
   \left(\frac{4\pi}{A_0}\right)^{2(\delta-1)}\tilde{r}_{A}^{4\delta-1}\right]\dot{\tilde{r}}_Adt.
\end{align}
Using Eqs. (\ref{4}), (\ref{8}), (\ref{11}), and (\ref{12}) in the first law of thermodynamics Eq. (\ref{9}), after some simplifications we find the Friedmann equation in differential form as
\begin{equation}\label{13}
  H(\rho+p)dt=\frac{1}{4\pi}\left[\frac{2\delta\pi}{G}\left(\frac{4\pi}{A_0}\right)^{\delta-1}\frac{1}{\tilde{r}_{A}^{5-2\delta}}
  +2(R-\delta)\frac{\delta\pi^2}{G^2}\left(\frac{4\pi}{A_0}\right)^{2(\delta-1)}\frac{1}{\tilde{r}_{A}^{5-4\delta}}\right]\dot{\tilde{r}}_Adt,
\end{equation}
which gives
\begin{equation}\label{14}
  \frac{1}{3}d\rho=-\left[\frac{\delta}{2G}\left(\frac{4\pi}{A_0}\right)^{\delta-1}\frac{d\tilde{r}_A}{\tilde{r}_{A}^{5-2\delta}}+
  (R-\delta)\frac{\delta\pi}{G^2}\left(\frac{4\pi}{A_0}\right)^{2(\delta-1)}\frac{d\tilde{r}_A}{\tilde{r}_{A}^{5-4\delta}}\right].
\end{equation}
Integration yeilds
\begin{equation}\label{15}
  \frac{1}{3}\int{d\rho}=-\left[\frac{\delta}{2G}\left(\frac{4\pi}{A_0}\right)^{\delta-1}\int{\frac{d\tilde{r}_A}{\tilde{r}_{A}^{5-2\delta}}}+
  (R-\delta)\frac{\delta\pi}{G^2}\left(\frac{4\pi}{A_0}\right)^{2(\delta-1)}\int{\frac{d\tilde{r}_A}{\tilde{r}_{A}^{5-4\delta}}}\right].
\end{equation}
Which gives
\begin{equation}\label{16}
  \frac{1}{3}\rho=\frac{\delta}{2G(4-2\delta)}\left(\frac{4\pi}{A_0}\right)^{\delta-1}\frac{1}{\tilde{r}_{A}^{4-2\delta}}+
  \frac{\delta\pi(R-\delta)}{2G^2(2-\delta)}\left(\frac{4\pi}{A_0}\right)^{2(\delta-1)}\frac{1}{\tilde{r}_{A}^{4(1-\delta)}},
\end{equation}
where the integration constant is set to be zero. Substituting $\tilde{r}_A$ from Eq. (\ref{3}), we obtain
\begin{equation}\label{17}
 \frac{1}{3}\rho=\frac{\delta}{2G(4-2\delta)}\left(\frac{4\pi}{A_0}\right)^{\delta-1}\left(H^2+\frac{k}{a^2}\right)^{2-\delta}+
  \frac{\delta\pi(R-\delta)}{2G^2(2-\delta)}\left(\frac{4\pi}{A_0}\right)^{2(\delta-1)}\left(H^2+\frac{k}{a^2}\right)^{2(\delta-1)}.
\end{equation}
The above equation can be written as
\begin{equation}\label{18}
  \left(H^2+\frac{k}{a^2}\right)^{2-\delta}+\frac{2\pi(R-\delta)}{G}\left(H^2+\frac{k}{a^2}\right)^{2(1-\delta)}\left(\frac{4\pi}{A_0}\right)^{\delta-1}=
  \frac{8\pi G_{eff}}{3}\rho,
\end{equation}
where
\begin{equation}\label{19}
  \frac{1}{G_{eff}}=\frac{2\delta}{G(2-\delta)}\left(\frac{4\pi}{A_0}\right)^{\delta-1}.
\end{equation}
The modified Friedmann equation based on the Sharma-Mittal entropy is found in equation (\ref{18}). In order to do this, we started with the first law of thermodynamics at the apparent horizon of a FRW universe and assume that the apparent horizon area has fractal features as a result of quantum-gravitational effects. Then, we derived the corresponding modified Friedmann equation for the FRW universe with any spatial curvature. When we put $\delta,R=1$, we get ($G_{eff}\rightarrow G$) the standard Friedmann equation in Einstein gravity.

Now we will compute the second Friedmann equation by combining first Friedmann equation (\ref{18}) with Eq. (\ref{6}). Taking the derivative of first Friedmann equation (\ref{18}), we get
\begin{equation}\label{20}
 \resizebox{.9\hsize}{!}{$ 2H\left(H^2+\frac{k}{a^2}\right)^{1-\delta}\left(\dot{H}+\frac{k}{a}\right)\left[(2-\delta)+\frac{2\pi(1-\delta)(R-\delta)}{G}
  \left(\frac{4\pi}{A_0}\right)^{\delta-1}\left(H^2+\frac{k}{a^2}\right)^{-\delta}\right]=\frac{8\pi G_{eff}}{3}\rho $}.
\end{equation}
Inserting Eq. (\ref{6}) in Eq. (\ref{20}), we get
\begin{equation}\label{21}
  \resizebox{.9\hsize}{!}{$ 2H\left(H^2+\frac{k}{a^2}\right)^{1-\delta}\left(\dot{H}+\frac{k}{a}\right)\left[(2-\delta)+\frac{2\pi(1-\delta)(R-\delta)}{G}
  \left(\frac{4\pi}{A_0}\right)^{\delta-1}\left(H^2+\frac{k}{a^2}\right)^{-\delta}\right]=-4G_{eff}(\rho+p) $}.
\end{equation}
Now using the relation $\dot{H}=\ddot{a}/a -H^2$, and $\rho$ from Eq. (\ref{18}), we can write the Eq. (\ref{21}) as
\begin{align}\label{22}
\nonumber
  &\left(H^2+\frac{k}{a^2}\right)^{1-\delta}\left(\frac{\ddot{a}}{a}-H^2-\frac{k}{a^2}\right)\left[(2-\delta)+\frac{2\pi(1-\delta)(R-\delta)}{G}
  \left(\frac{4\pi}{A_0}\right)^{\delta-1}\left(H^2+\frac{k}{a^2}\right)^{-\delta}\right] \\
   &=-4\pi G_{eff}p-\frac{3}{2}\left(H^2+\frac{k}{a^2}\right)^{2-\delta}+\frac{\pi(R-\delta)}{G}\left(H^2+\frac{k}{a^2}\right)^{2(1-\delta)}
   \left(\frac{4\pi}{A_0}\right)^{\delta-1}.
\end{align}
After simplification, we obtain
\begin{align}\label{23}
\nonumber
  &\resizebox{1.0\hsize}{!}{$(2-\delta)\left(H^2+\frac{k}{a^2}\right)^{1-\delta}\frac{\ddot{a}}{a}+(2+\delta)\left(H^2+\frac{k}{a^2}\right)^{2-\delta}+
  \frac{2\pi(1-\delta)(R-\delta)}{G}\left(\frac{4\pi}{A_0}\right)^{\delta-1}\left(H^2+\frac{k}{a^2}\right)^{1-2\delta}\frac{\ddot{a}}{a} $}\\
  &+(\delta+2)\frac{2\pi(R-\delta)}{G}\left(\frac{4\pi}{A_0}\right)^{\delta-1}\left(H^2+\frac{k}{a^2}\right)^{2(1-\delta)}=-8\pi G_{eff}p.
\end{align}
This is the second Friedmann equation governing the evolution of the universe based on Shirma-Mittal entropy. The equation (\ref{23}) reproduces the second Friedmann equation in standard cosmology, which has the following form for $R, \delta=1$ ($G_{eff}\rightarrow G$)
\begin{equation}\label{24}
  \frac{\ddot{a}}{a}+3\left(H^2+\frac{k}{a^2}\right)=-8\pi Gp.
\end{equation}
If we combine the Eq. (\ref{18}) and Eq. (\ref{23}), we get
\begin{align}\label{25}
\nonumber
  &(2-\delta)\left(H^2+\frac{k}{a^2}\right)^{1-\delta}\frac{\ddot{a}}{a}+
   \frac{2\pi(1-\delta)(R-\delta)}{G}\left(\frac{4\pi}{A_0}\right)^{\delta-1}\left(H^2+\frac{k}{a^2}\right)^{1-2\delta}\frac{\ddot{a}}{a}  \\
   &=-\frac{8\pi G_{eff}}{3}\rho\left[(2+\delta)+3p\right],
\end{align}
where $\omega=p/\rho$ is the equation of state parameter. Taking into account the fact that $0\leq\delta\leq1, 0<R\leq1$, the condition for the accelerated expansion of the universe ($\ddot{a}>0$), implies
\begin{equation}\label{26}
  (2+\delta)+3\omega<0\quad\longrightarrow \quad\omega<-\frac{2+\delta}{3}.
\end{equation}
For $\delta, R=0$, it suggests the convoluted and fractal structure, so that $\omega<-2/3$ which also corresponds the horizon structure with area law, while for $\delta, R=1$ which corresponds to the more intricate feature and fractal structure, so that $\omega<-1$. This suggests that, the equation of state parameter is forced to grow increasingly negative in an accelerating universe due to the fractal structure of the apparent horizon.

This segment summarizes that the Sharma-Mittal cosmology's modified equations, namely Eqs. (\ref{18}) and (\ref{25}), can illustrate the universe's evolution with any spatial curvature when the entropy connected to the visible horizon transforms due to quantum-gravitational consequences. The cosmological implications of the Friedmann equations derived above are left for further research.

\section{Generalized Second Law of Thermodynamics}
In this section, we have established the generalized second law of thermodynamics by utilizing the SM entropy (\ref{1}) which is linked to the fractal arrangement of the universe's horizon area, by taking into account the apparent horizon that encompasses the universe. It differs from the one presented in \cite{E34} where the authors modified the total energy density in the Friedmann equations by using the Barrow entropy. However, in a flat universe with a dark energy sector acting as an additional energy component. The cosmological field equations provided in equations (\ref{3}) and (\ref{4}) of \cite{E34} reduce to the standard Friedmann equations \cite{E60}. Thus, we assume that the SM entropy does not impact the energy content of the universe. Furthermore, unlike the authors of \cite{E34}, who considered only a flat universe, we consider any specific curvature of the FRW universe. The generalized second law of thermodynamics has been examined in the context of the expanding universe in \cite{E61,E62,E63}.

Combining Eq. (\ref{14}) with Eq. (\ref{6}) and using Eq. (\ref{19}), we get
\begin{equation}\label{27}
  \frac{2(2-\delta)}{\tilde{r}_{A}^{5-2\delta}}\dot{\tilde{r}}_{A}+\frac{2\pi\delta(2-\delta)(R-\delta)}{\tilde{r}_{A}^{5-4\delta}}
  \left(\frac{4\pi}{A_0}\right)^{\delta-1}\dot{\tilde{r}}_{A}=8\pi G_{eff}H(\rho+p).
\end{equation}
Solving for $\dot{\tilde{r}}_{A}$, we have
\begin{equation}\label{28}
 \dot{\tilde{r}}_{A}=\frac{8\pi G_{eff}}{2(2-\delta)}H(\rho+p)\tilde{r}^{5-2\delta}_{A}+\frac{8\pi^2 G_{eff}(\delta-R)}{2(2-\delta)}
  \left(\frac{4\pi}{A_0}\right)^{\delta-1}\tilde{r}_{A}^{5}H(\rho+p).
\end{equation}
Since $\delta, R\leq1$, thus sign of $\dot{\tilde{r}}_A$ depends on the sign of ($\rho+p$). In case when $\rho+p\geq0$, the dominant energy condition(DEC)is satisfied then we have $\dot{\tilde{r}}_A\geq0$. Now we calculate $T_h\dot{S}_{h}^{SM}$, as
\begin{align}\label{29}
\nonumber
 &T_h\dot{S}_{h}^{SM}=\\
 &\frac{\delta}{G}\left(1-\frac{\dot{\tilde{r}}_A}{2H\tilde{r}_A}\right) \left(\frac{4\pi}{A_0}\right)^{\delta-1}\tilde{r}_{A}^{2(\delta-1)}\dot{\tilde{r}}_A+
 \frac{\delta\pi(R-\delta)}{G}\left(1-\frac{\dot{\tilde{r}}_A}{2H\tilde{r}_A}\right) \left(\frac{4\pi}{A_0}\right)^{2(\delta-1)}\tilde{r}_{A}^{2(2\delta-1)}\dot{\tilde{r}}_A.
\end{align}
Substituting Eqs. (\ref{19}) and (\ref{28}) in Eq. (\ref{29}), we get
\begin{align}\label{30}
\nonumber
&T_h\dot{S}_{h}^{SM}=2\pi H\tilde{r}_{A}^{3}(\rho+p)\left(1-\frac{\dot{\tilde{r}}_A}{2H\tilde{r}_A}\right)+  \\
&2\pi^2 H(R-\delta)(\rho+p)\tilde{r}_{A}^{3+2\delta}(\rho+p)\left(1-\frac{\dot{\tilde{r}}_A}{2H\tilde{r}_A}\right)
\left[\left(\frac{4\pi}{A_0}\right)^{\delta-1}-1+\frac{\delta-R}{G}\left(\frac{4\pi}{A_0}\right)^{\delta-1}\tilde{r}^{2\delta}_{A}\right].
\end{align}
When we set $R, \delta=1$ we get the following form
\begin{equation*}
  T_h\dot{S}_{h}^{SM}=2\pi H\tilde{r}_{A}^{3}(\rho+p)\left(1-\frac{\dot{\tilde{r}}_A}{2H\tilde{r}_A}\right).
\end{equation*}
For accelerating universe, the equation of state parameter can cross the photon line ($\omega=p/\rho <-1$), which means that the DEC may violate, $\rho+p<0$. As the result, when we use $R=\delta$ and $\rho+p<0$, the DEC valid everywhere, $\dot{S}_{h}^{SM}\geq0$ is valid everywhere. In the second case, we use $\delta$ less than $R$ and $\rho+p<0$ implying that the second law of thermodynamics may no longer valid. So for this, we can assume the total entropy of the universe as $S_{\text{total}}= S^{SM}_{h}+S_m$, where $S_m$ is the matter entropy inside the apparent horizon. If the generalized second law of thermodynamics holds, we should have $\dot{S}_{SM}^{h}+\dot{S}_m\geq0$, for the total entropy.

From the Gibbs equation we have \cite{E64}
\begin{equation}\label{31}
  T_m dS_m=d(\rho V)+p dV=Vd\rho+(\rho+p)dV,
\end{equation}
where $T_m$ denotes the temperature of the matter field within the apparent horizon, which we propose in thermal equilibrium with the surrounding universe. This equilibrium hypothesis, as outlined in reference \cite{E65}, implies that the temperature of the matter field within the universe is uniform and is equal to the temperature of the apparent horizon boundary, i.e., $T_m = T_h$. Without this local equilibrium hypothesis, there would be spontaneous heat transfer between the horizon and the bulk fluid, which is not feasible in our universe from a physical standpoint. So, from Eq. (\ref{31}), we obtain
\begin{equation}\label{32}
  T_m \dot{S}_m=4\pi\tilde{r}_{A}^{2}\dot{\tilde{r}}_A-4\pi\tilde{r}_{A}^{3}H(\rho+p).
\end{equation}
Beside this, we consider the generalised second law of thermodynamics, specifically how total entropy changes over time. For this, combining Eqs. (\ref{30}) and (\ref{32}), we obtain
\begin{align}\label{33}
\nonumber
&T_h(\dot{S}_{h}^{SM}+\dot{S}_m)=8\pi\tilde{r}_{A}^{2}(\rho+p)\dot{\tilde{r}}_A-2\pi\tilde{r}_{A}^{3}H(\rho+p)+ \\
 &2\pi^2(R-\delta)H\tilde{r}_{A}^{3+2\delta}\left(1-\frac{\dot{\tilde{r}}_A}{2H\tilde{r}_A}\right)\left[\left(\frac{4\pi}{A_0}\right)^{\delta-1}-1
 +\frac{\delta-R}{G}\left(\frac{4\pi}{A_0}\right)^{\delta-1}\tilde{r}^{2\delta}_{A}\right].
\end{align}
Using Eq. (\ref{28}) in Eq. (\ref{33}), we obtain
\begin{align}\label{34}
\nonumber
  &T_h(\dot{S}_{h}^{SM}+\dot{S}_m)=2\pi H(\rho+p)\tilde{r}_{A}^{3}\left[\left(\left(\frac{4\pi}{A_0}\right)^{\delta-1}-1\right)\tilde{r}_{A}^{2\delta}+\frac{\delta-R}{G}\tilde{r}_{A}^{4\delta}
  -1\right]+  \\
&\resizebox{1.0\hsize}{!}{$\frac{16\pi^2}{2-\delta}G_{eff}H(\rho+p)^2\tilde{r}_{A}^{7-2\delta}\left[1+\pi(R-\delta)\left(\frac{4\pi}{A_0}\right)^{\delta-1}
\tilde{r}_{A}^{2\delta}+\left(\left(\frac{4\pi}{A_0}\right)^{\delta-1}-1+\frac{R-\delta}{G}\left(\frac{4\pi}{A_0}\right)^{\delta-1}
\tilde{r}_{A}^{2\delta}\right)(1+\pi\tilde{r}_{A}^{2\delta})\frac{\pi(\delta-R)}{4}\tilde{r}_{A}^{2\delta}\right]$},
\end{align}
which throughout the duration of the universe has had a non-negative function. The DEC is also satisfied; demonstrates that the generalized second law of thermodynamics applies to universes with fractal boundaries, specifically when the associated entropy with the universe's apparent horizon takes the form of Sharma-Mittal entropy. Furthermore, when $\delta\rightarrow R$, the result for Tsallis entropy is also recovered, In this case, we obtain
\begin{equation}\label{35}
  T_h(\dot{S}_T+\dot{S}_m=\frac{16\pi^2}{2-\delta}G_{eff}H(\rho+p)\tilde{r}_{A}^{7-2\delta}+2\pi H(\rho+p)\tilde{r}_{A}^{3}
  \left[\left(\left(\frac{4\pi}{A_0}\right)^{\delta-1}-1\right)\tilde{r}_{A}^{2\delta}-1\right].
\end{equation}
It also has a non-negative function throughout the history of the universe. So, it is clear that the generalized second law of thermodynamics is valid everywhere in both cases (Tsallis and Sharma-Mittal entropy for the universe with the fractal boundary) by using the apparent horizon of the universe.

\section{Modified Friedmann Equation from Emergence of Cosmic Space}
In this section, we have adopted the perspectives of Padmanabhan \cite{E36} and Verlinde \cite{E41} on gravity. Verlinde argues that gravity is a force that arises from emergence, while Padmanabhan employs the equipartition law to explain the universe's expansion, using the concept of the holographic principle. Padmanabhan asserts that the difference between the degrees of freedom on the holographic surface and those in the emerging bulk is directly proportional to the alteration in the cosmic volume, successfully recovering the Friedmann equation \cite{E51}. This equation governs the universe's development in the absence of spatial curvature. From this standpoint, cosmic space emerges and advances with cosmic time, and our universe's spatial expansion results from the emergence of space.. According to Padmanabhan's idea, the increase of cosmic volume $dV$ in an infinitesimal time $dt$ of cosmic time is given by
\begin{equation}\label{36}
  \frac{dV}{dt}=G(N_\text{sur}-N_\text{bulk}).
\end{equation}
Here, $N_\text{sur}$ and $N_\text{bulk}$ are the degrees of freedom on the boundary and in the bulk, respectively. Following Padmanabhan, the studies were generalized to Gauss-Bonnet and Lovelock gravity \cite{E52}. The authors in \cite{E52} derived the Friedmann equations for Einstein gravity with varying spatial curvatures, but they were unable to apply this method to higher-order gravity theories such as Gauss-Bonnet and Lovelock gravities in the case of non-flat FRW universes. In \cite{E54}, the author modified Padmanabhan's approach to obtain the Friedmann equations for higher-order gravity theories with different spatial curvatures. The modified form of Eq. (\ref{36}), is given by
\begin{equation}\label{37}
  \frac{dV}{dt}=G\frac{\tilde{r}_A}{H^{-1}}(N_\text{sur}-N_\text{bulk}).
\end{equation}
In contrast to Padmanabhan's original equation (\ref{36}), the volume increase in a non-flat universe is still linked to the difference between degrees of freedom on the apparent horizon and the bulk, but the proportionality factor is now the ratio of the apparent horizon to the Hubble radius, and it is not a constant. If the universe is flat, this ratio becomes equal to 1, and Padmanabhan's original equation (\ref{36}) is recoverd. Our aim is to develop a new Friedmann equation based on Sharma-Mittal entropy and the concept of cosmic space emergence. Using Sharma-Mittal's entropy expression (\ref{1}) as a guide, we define the effective area of the apparent horizon, which serves as holographic boundary, as
\begin{equation}\label{38}
  \tilde{A}=A^{\delta}=(4\pi\tilde{r}_{A}^{2})^\delta.
\end{equation}
Next, we determine the increase in the effective volume as
\begin{align}\label{39}
\nonumber
  \frac{d\tilde{V}}{dt}&=\frac{\tilde{r}_A}{2}\frac{d\tilde{A}}{dt}=\delta(4\pi\tilde{r}_{A}^{2})^\delta\dot{\tilde{r}}_A  \\
  &=\frac{\delta}{2(\delta-2)}(4\pi)^\delta(\tilde{r}_{A}^{5})\frac{d}{dt}\left(\tilde{r}_{A}^{2(\delta-2)}\right).
\end{align}
Following [36] and from Eq. (\ref{39}), we take
\begin{equation}\label{40}
  N_{\text{sur}}=\frac{4\pi\tilde{r}_{A}^{2\delta}}{G_{eff}}.
\end{equation}
We have utilized (\ref{19}) and assumed that the temperature of the apparent horizon is the Hawking temperature, indicated in \cite{E52} as
\begin{equation}\label{41}
  T=\frac{1}{2\pi\tilde{r}_A}.
\end{equation}
The energy inside a sphere with volume $V=\frac{4\pi}{3}\tilde{r}_{A}^{3}$ is referred to as the Komar energy
\begin{equation}\label{42}
  E_{\text{komar}}=|(\rho+3p)|V.
\end{equation}
By using the equipartition law of energy, the bulk degrees of freedom can be defined as
\begin{equation}\label{43}
  N_{\text{bulk}}=\frac{2|E_{\text{komar}}|}{T}.
\end{equation}
We take $\rho+3p<0$ in order to have $N_{\text{bulk}}>0$ [33]. Consequently, the bulk's degree of freedom is calculated as
\begin{equation}\label{44}
  N_{\text{bulk}}=-\frac{16\pi^2}{3}\tilde{r}_{A}^{4}(\rho+3p).
\end{equation}
Using the appropriate equation (\ref{37}) is the second key assumption in this situation. We use, $G\rightarrow \Gamma^{-1}$ and $V\rightarrow\tilde{V}$ in order to produce the proper form in Eq. (\ref{37}), which can be rewritten as
\begin{equation}\label{45}
  \Gamma\frac{d\tilde{V}}{dt}=\frac{\tilde{r}_A}{H^{-1}}(N_\text{sur}-N_\text{bulk}),
\end{equation}
where $\Gamma=A_0/G$. Substituting Eqs. (\ref{39}), (\ref{40}) and (\ref{44}) in Eq. (\ref{45}), we obtain
\begin{equation}\label{46}
  \frac{A_0}{G}\delta(4\pi)^\delta\tilde{r}_{A}^{2\delta}=\frac{\tilde{r}_A}{H^{-1}}\left[\frac{4\pi\tilde{r}_{A}^{2\delta}}{G_{eff}}+
  \frac{16\pi^2}{3}\tilde{r}_{A}^{4}(\rho+3p)\right].
\end{equation}
Using Eq. (\ref{19}) in Eq. (\ref{46}), we can write
\begin{equation}\label{47}
  (2-\delta)\tilde{r}_{A}^{2\delta-5}\frac{\dot{\tilde{r}}_A}{H}-\tilde{r}_{A}^{2(\delta-2)}=\frac{8\pi G_{eff}}{3}(\rho+3p).
\end{equation}
Multiplying by $2\dot{a}a$ on both side and using Eq. (\ref{6}), we get
\begin{equation}\label{48}
  \frac{d}{dt}\left(a^2\tilde{r}_{A}^{2(\delta-2)}\right)=\frac{8\pi G_{eff}}{3}\frac{d}{dt}(\rho a^2).
\end{equation}
Integrating above equation and using the Eq. (\ref{3}), we get
\begin{equation}\label{49}
  \left(H^2+\frac{k}{a^2}\right)^{2-\delta}=\frac{8\pi G_{eff}}{3}\rho.
\end{equation}
Our research confirms the legitimacy of Padmanabhan's modified perspective on gravity as an emergent force, as indicated by Eq. (\ref{45}). We observed that the adjusted Friedmann equation, derived from the emergent approach, matches the one derived from the first law of thermodynamics. This emphasizes the effectiveness of the method presented in capturing the emergence of cosmic space when the apparent horizon's associated entropy complies with SM entropy (\ref{1}).
\section{Conclusions}
Recently, Barrow proposed a new expression for black hole entropy, inspired by the structure of the COVID-19 virus \cite{E1}. The author showed that considering quantum-gravitational effects can result in fractal, complex features of the black hole horizon. This structure has a finite volume but an infinite or finite area. As a result, the entropy of the black hole horizon, altered by the quantum-gravitational effects, deviates from the area law and increases. The increase in entropy is proportional to the quantum-gravitational deformation of the horizon, quantified by the exponent $\delta$ and $R$.

We investigated the corrections to the Friedmann equations of an FRW universe with any spatial curvature, based on the assumption that the entropy of the apparent horizon of the universe follows the same expression as black hole entropy (Sharma-Mittal proposal). These corrections are caused by the quantum-gravitational, fractal and complex structure of the apparent horizon.

We examined the "gravity-thermodynamics" hypothesis by proposing that the first law of thermodynamics ($dE = T_{h}dS^{SM}_{h} + WdV$) is applicable to the apparent horizon of the FRW universe, with SM entropy (\ref{1}) as the entropy measure. By employing the first law and SM entropy, we derived modified Friedmann equations to describe the universe's evolution. We than evaluated the generalized second law of thermodynamics by analyzing the time-dependent behavior of both matter entropy and the SM entropy of the apparent horizon. Furthermore, we implemented Padmanabhan's concept of emergent gravity and determined the degrees of freedom in the bulk and boundary of the universe. We obtained modified Friedmann equations by subtracting the number of degrees of freedom in the boundary and bulk and utilizing Padmanabhan's revised proposal in Eq. (\ref{45}), which accounts for SM entropy. This outcome is consistent with the findings derived from the first law of thermodynamics. Our study corroborates the validity of the emergence gravity theory proposed in references \cite{E51,E54}.

There are multiple areas that warrant further exploration, such as the consequences of the modified Friedmann equations on the universe's evolution, gravitational collapse, structure formation, and galaxy development. Another fascinating subject is the influence of the fractal parameter $\delta$ on the thermal history and CMB anisotropy. Nevertheless, these topics extend beyond the scope of this study and will be presented for future investigation.
\section*{Authors Contribution}
MN: Conceptualization, software, writing the original draft, formal analysis. AB: Conceptualization, formal analysis and validation.
\section*{Funding}
Not applicable.
\section*{Data Availability Statement}
No data was used for the research described in the article.
\section*{Declarations}
\textbf{Conflict of interest:} The authors declare that they have no competing interest.

\end{document}